\documentclass{article}

\usepackage{spconf}
\usepackage{amsmath, amsfonts, graphicx, booktabs, tabularx, multirow, siunitx, xcolor, soul, bm, enumitem, tikz, mathtools}
\usepackage{hyperref}

\graphicspath{{./figures/}}

\title{Image-Based Soil Organic Carbon Remote Sensing from Satellite Images with Fourier Neural Operator and Structural Similarity}
\name{
\begin{tabular}{c}
Ken C. L. Wong$^{1,*}$, Levente Klein$^2$, Ademir Ferreira da Silva$^3$, \\Hongzhi Wang$^1$, Jitendra Singh$^4$, Tanveer Syeda-Mahmood$^1$
\end{tabular}
\thanks{This paper was accepted by IEEE IGARSS 2023 (\url{https://doi.org/10.1109/IGARSS52108.2023.10281551}). \copyright 2023 IEEE. Personal use of this material is permitted. Permission from IEEE must be obtained for all other uses, in any current or future media, including reprinting/republishing this material for advertising or promotional purposes, creating new collective works, for resale or redistribution to servers or lists, or reuse of any copyrighted component of this work in other works.
}
\thanks{
\noindent$^{*}$Corresponding author (clwong@us.ibm.com).
}
}
\address{
$^1$IBM Research -- Almaden Research Center, San Jose, CA, USA \\
$^2$IBM Thomas J. Watson Research Center, Yorktown Heights, NY, USA \\
$^3$IBM Research, Rio de Janeiro, RJ, Brazil \\
$^4$IBM Research, Gurgaon, India
}
\begin{document}
%\ninept
%
\maketitle
\begin{abstract}
  Soil organic carbon (SOC) sequestration is the transfer and storage of atmospheric carbon dioxide in soils, which plays an important role in climate change mitigation. SOC concentration can be improved by proper land use, thus it is beneficial if SOC can be estimated at a regional or global scale. As multispectral satellite data can provide SOC-related information such as vegetation and soil properties at a global scale, estimation of SOC through satellite data has been explored as an alternative to manual soil sampling. Although existing studies show promising results, they are mainly based on pixel-based approaches with traditional machine learning methods, and convolutional neural networks (CNNs) are uncommon. To study the use of CNNs on SOC remote sensing, here we propose the FNO-DenseNet based on the Fourier neural operator (FNO). By combining the advantages of the FNO and DenseNet, the FNO-DenseNet outperformed the FNO in our experiments with hundreds of times fewer parameters. The FNO-DenseNet also outperformed a pixel-based random forest by 18\% in the mean absolute percentage error.
\end{abstract}
\section{Introduction}

Carbon sequestration is the process of transferring atmospheric carbon dioxide (CO$_2$), which is the main cause of global warming, into a carbon pool and storing it securely \cite{Journal:Lal:Science2004:soil}. Atmospheric CO$_2$ can be transferred and stored in soil in the form of soil organic carbon (SOC), and the effectiveness of SOC sequestration can be improved by methods such as forest management and conservation tillage of croplands. As SOC plays an important role in climate change mitigation \cite{Journal:Bossio:Nature2020:current}, it is beneficial if the SOC level can be accurately and efficiently estimated to allow monitoring and management.

Estimating SOC at a global scale is challenging due to variability in soils, complex biochemical processes, and different management practices \cite{Journal:Poggio:Soil2021:soilgrids}. Traditionally, soil is sampled at different depths and lab analyzed to measure SOC, but manual sampling is impractical at a global scale. In contrast, satellite data provide a complementary, scalable, and cost-effective alternative. Remote sensing through satellites has the potential to enable measurement, reporting, and verification of SOC across the globe, with year-to-year tracking of carbon storage and carbon cycle disruptions when carbon sequestration practices are implemented \cite{Journal:Smith:GCB2020:current}.

As multispectral and hyperspectral satellite data can provide SOC related information such as vegetation, water, and soil properties, prediction and mapping of SOC using these data is an active research topic \cite{Journal:Vzivzala:RS2019:soil,Journal:Angelopoulou:RS2019:remote,Journal:Meng:AEOG2020:regional,Journal:Venter:STTE2021:mapping,Journal:Wang:RSE2022:using}. Although the results are promising, most studies use pixel-based regression with traditional machine learning methods such as support vector machines and random forests. Deep learning approaches, especially convolutional neural networks (CNNs), are seldom used \cite{Journal:Meng:AEOG2020:regional,Journal:Wang:RSE2022:using}. In this paper, to study the advantages of using CNNs on SOC remote sensing, we propose an image-based deep learning approach to estimate SOC from multispectral satellite imagery. By modifying the state-of-the-art Fourier neural operator (FNO) \cite{Conference:Li:ICLR2021:fourier}, we propose the FNO-DenseNet by combining the advantages of the FNO and DenseNet \cite{Conference:Huang:CVPR2017:densely}. We also found that incorporating structural similarity (SSIM) \cite{Journal:Wang:TIP2004:image} in the loss function can further improve SOC estimation. In our experiments, the FNO-DenseNet outperformed the FNO and had hundreds of times fewer parameters. The FNO-DenseNet also outperformed a modified V-Net \cite{Conference:Wong:MICCAI2018} and a pixel-based random forest.

\section{Methodology}

% ------------------ figure -------------------------
\begin{figure}[t]
    \centering
    \includegraphics[width=1\linewidth]{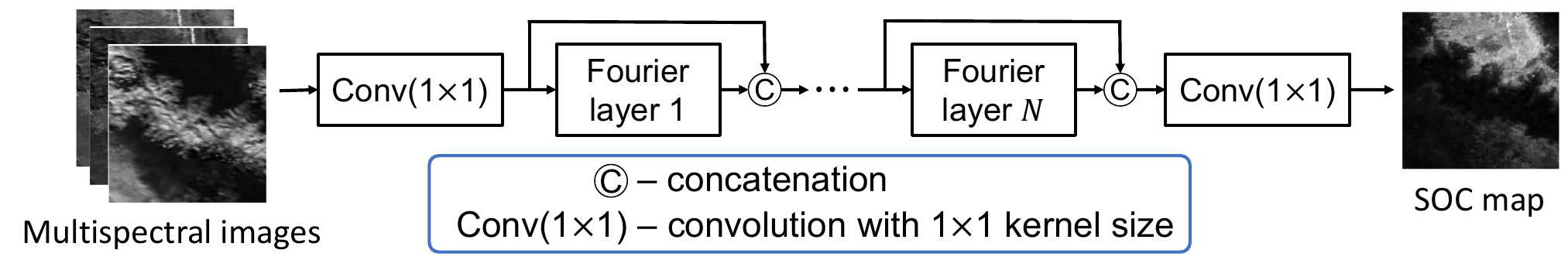}
    \caption{FNO-DenseNet. Multispectral images are stacked as multi-channel inputs. The Fourier layers are (\ref{eq:fno_update}) implemented by the fast Fourier transform. $N=8$ was used in the experiments, and all Conv and Fourier layers had 24 output channels. For the original FNO, we used $N=8$ and 32 output channels without concatenations.}
    \label{fig:network}
\end{figure}
% ---------------------------------------

\subsection{FNO-DenseNet for Image-Based SOC Estimation}

The FNO is a deep learning model that was proposed to learn mappings between functions in partial differential equations (PDEs) \cite{Conference:Li:ICLR2021:fourier}. As the formulations were developed based on the Green's function in the continuous space, the model training does not require the actual PDEs, and the trained model is theoretically independent of data resolution and discretization \cite{Journal:Li:arXiv2020:neural}. Therefore, it is beneficial to use the FNO to learn the underlying relation between the multispectral images and SOC data. The neural operator is formulated as iterative updates:
\begin{equation}\label{eq:fno_update}
    \begin{gathered}
      v_{t+1}(x) \coloneqq \sigma \left(W v_t(x) + \left(\mathcal{K}v_t\right)(x)\right) \\
      \textrm{with} \ \ \left(\mathcal{K}v_t\right)(x) \coloneqq \int_{D} \kappa(x - y)v_t(y) \,dy, \ \ \forall x \in D
    \end{gathered}
\end{equation}
where $v_t(x) \in \mathbb{R}^{d_{v_t}}$ is a function and $W \in \mathbb{R}^{d_{v_{t+1}} \times d_{v_t}}$ is a learnable linear transformation. $\sigma$ accounts for normalization and activation. In our work, $D \subset \mathbb{R}^2$ represents the 2D imaging space, and $v_t(x)$ are the outputs of hidden layers with $d_{v_t}$ channels. $\mathcal{K}$ is the kernel integral operator with $\kappa \in \mathbb{R}^{d_{v_{t+1}} \times d_{v_t}}$ a learnable kernel function. As $\left(\mathcal{K}v_t\right)(x)$ is a convolution operator, it can be efficiently computed using the convolution theorem with the Fourier transform $\mathcal{F}$:
\begin{equation}\label{eq:fourier_conv}
    \begin{split}
      \left(\mathcal{K}v_t\right)(x) &= \mathcal{F}^{-1}\left(\mathcal{F}(\kappa) \cdot \mathcal{F}(v_t)\right)(x) \\
      &= \mathcal{F}^{-1}\left(R \cdot \left(\mathcal{F}v_t\right)\right)(x), \ \ \forall x \in D
    \end{split}
\end{equation}
For each $k$ in the frequency domain, $R(k) \in \mathbb{C}^{d_{v_{t+1}} \times d_{v_t}}$ is learnable, and $\left(\mathcal{F}v_t\right)(k) \in \mathbb{C}^{d_{v_t}}$. The Fourier transform also provides a global receptive field as all pixels are used to compute the value at each $k$ during implementation.

To improve feature reuse and propagation, following DenseNet \cite{Conference:Huang:CVPR2017:densely}, we modify the original FNO by concatenating $v_t(x)$ (Fig. \ref{fig:network}). Moreover, instead of using a different $R(k)$ for each $k$, we find that using a shared $R$ does not diminish the accuracy while the number of parameters can be significantly reduced (e.g., by more than 500 times).

\subsection{Loss Function with Structural Similarity}

In our early experiments, we used the mean absolute error (MAE) as the loss function as it outperformed the mean squared error. Nevertheless, we found that using the MAE alone produced blurry predictions that lacking the structural details of the ground truths. This is most likely due to the texture-like appearances of the SOC data and the pixel-based MAE cannot account for the local dependencies among pixels. Therefore, SSIM is used \cite{Journal:Wang:TIP2004:image}, whose value ($\in$ [-1, 1]) between two nonnegative images $\mathbf{x}$ and $\mathbf{y}$ is:
\begin{equation}\label{eq:ssim_window}
  \mathrm{SSIM}(\mathbf{x}, \mathbf{y}) = \frac{1}{M} \sum_{j=1}^{M}\frac{\left(2\mu_x\mu_y + C_1\right)\left(2\sigma_{xy} + C_2\right)}{\left(\mu_x^2 + \mu_y^2 + C_1\right)\left(\sigma_x^2 + \sigma_y^2 + C_2\right)}
\end{equation}
where $C_1$ and $C_2$ are small constants to avoid divide by zero. The means ($\mu_x$, $\mu_y$), variances ($\sigma_x^2$, $\sigma_y^2$), and covariance ($\sigma_{xy}$) are computed from a local window $j$ of size 11$\times$11 pixels using a Gaussian weighting function with standard deviation of 1.5 pixels. The SSIM compares the luminance, contrast, and structure between $\mathbf{x}$ and $\mathbf{y}$. Combining with the MAE, the overall loss function is given as:
\begin{equation}\label{eq:loss}
  L = w \times \mathrm{MAE} + 0.5 \times \left(1 - \mathrm{SSIM}\right) = w \times \mathrm{MAE} + \mathrm{DSSIM}
\end{equation}
where $\mathrm{DSSIM} = 0.5 \times \left(1 - \mathrm{SSIM}\right) \in [0, 1]$ is the structural dissimilarity. $w$ is a scalar accounting for the difference in magnitude, which is 0.01 in our experiments.

\subsection{Data and Training Strategy}
Six spectral bands from the Moderate Resolution Imaging Spectroradiometer (MODIS) \cite{Misc:MODIS}, i.e., blue (459–479 nm), green (545–565 nm), red (620–670 nm), near infrared (841–876 nm), and shortwave infrared (SWIR$_1$: 1230–1250 nm, SWIR$_3$: 2105–2155 nm), were used as the predictors. The SOC data in the top 5 cm from SoilGrids \cite{Journal:Poggio:Soil2021:soilgrids} were the predictands. Sampling from the regions of USA, Mexico, and Canada, a total of 3059 samples were generated. The dataset was partitioned into 50\% for training, 20\% for validation, and 30\% for testing, and identical partitions were used in all experiments. All images were resampled to 128$\times$128 pixels with a spatial resolution of 500 m. Each model was trained for 400 epochs with the batch size of 32. The Adamax optimizer \cite{Journal:Kingma:arXiv2014} was used with the cosine annealing learning rate scheduler \cite{Conference:Loshchilov:ICLR2017:SGDR}, with the maximum and minimum learning rates as $10^{-2}$ and $10^{-4}$, respectively. Image augmentation with rotation ($\pm$30$^{\circ}$), shifting ($\pm$20\%), scaling ($\in$ [0.8, 1.2]), and flipping were used with an 80\% chance.

% ------------------ Table -------------------------
\begin{table*}[t]
\caption{Testing results (mean$\pm$std) from five repeats on the same testing data. The unit of RMSE is g/kg, MAPE is shown in percentage, and SSIM is unitless. The best results of each column are highlighted. Each random forest model had 10 trees with a maximum depth of 10. The numbers of parameters of the modified V-Net, FNO, and FNO-DenseNet are 342K, 34M, and 64K, respectively. Note that the DSSIM loss cannot be applied to the pixel-based random forest.}
\label{table:results}

\smallskip
\fontsize{7}{8}\selectfont
\centering

\newcolumntype{C}{>{\centering\arraybackslash}X}
\newcommand{\boldblue}[1]{\textcolor{blue}{\textbf{#1}}}

\begin{tabularx}{\linewidth}{@{\extracolsep{4pt}}lCCCCCCCCC}
\toprule
Loss & \multicolumn{3}{c}{MAE} & \multicolumn{3}{c}{DSSIM} & \multicolumn{3}{c}{MAE + DSSIM}\\
\cline{1-1} \cline{2-4} \cline{5-7} \cline{8-10} \noalign{\smallskip}
Metric & \multicolumn{1}{c}{RMSE} & \multicolumn{1}{c}{MAPE} & \multicolumn{1}{c}{SSIM} & \multicolumn{1}{c}{RMSE} & \multicolumn{1}{c}{MAPE} & \multicolumn{1}{c}{SSIM} & \multicolumn{1}{c}{RMSE} & \multicolumn{1}{c}{MAPE} & \multicolumn{1}{c}{SSIM} \\
\midrule
Random forest & 2.50$\pm$2.37 & 45.40$\pm$46.30 & 0.07$\pm$0.08 & ------ & ------ & ------ & ------ & ------ & ------ \\
Modified V-Net & 2.12$\pm$2.26 & 30.12$\pm$23.00 & 0.07$\pm$0.06 & 2.21$\pm$2.25 & 33.44$\pm$22.23 & \boldblue{0.20$\pm$0.13} & 2.00$\pm$1.99 & 29.79$\pm$21.11 & 0.17$\pm$0.12 \\
FNO & \boldblue{1.97$\pm$2.03} & 28.08$\pm$21.13 & 0.10$\pm$0.09 & 2.31$\pm$2.34 & 35.57$\pm$24.22 & \boldblue{0.20$\pm$0.13} & 1.96$\pm$1.96 & 28.32$\pm$18.29 & \boldblue{0.18$\pm$0.13} \\
FNO-DenseNet & 1.98$\pm$2.09 & \boldblue{27.24$\pm$19.25} & \boldblue{0.11$\pm$0.10} & \boldblue{2.16$\pm$2.13} & \boldblue{32.33$\pm$19.48} & \boldblue{0.20$\pm$0.13} & \boldblue{1.89$\pm$1.75} & \boldblue{27.75$\pm$18.12} & \boldblue{0.18$\pm$0.12} \\
\bottomrule
\end{tabularx}
\end{table*}
% --------------------------------------------------

% ------------------ figure -------------------------
\begin{figure*}[t]
    %\scriptsize
    \centering
    \includegraphics[width=0.12\linewidth]{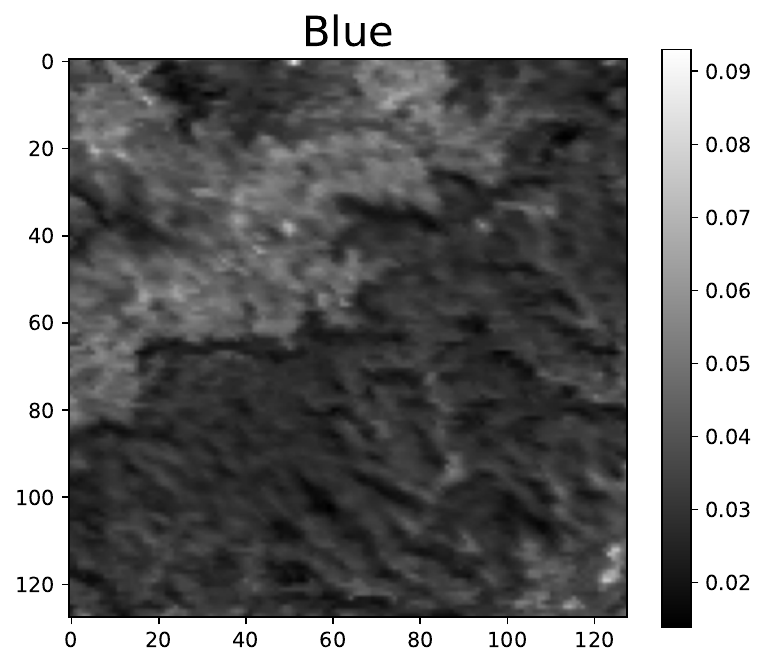}
    \includegraphics[width=0.12\linewidth]{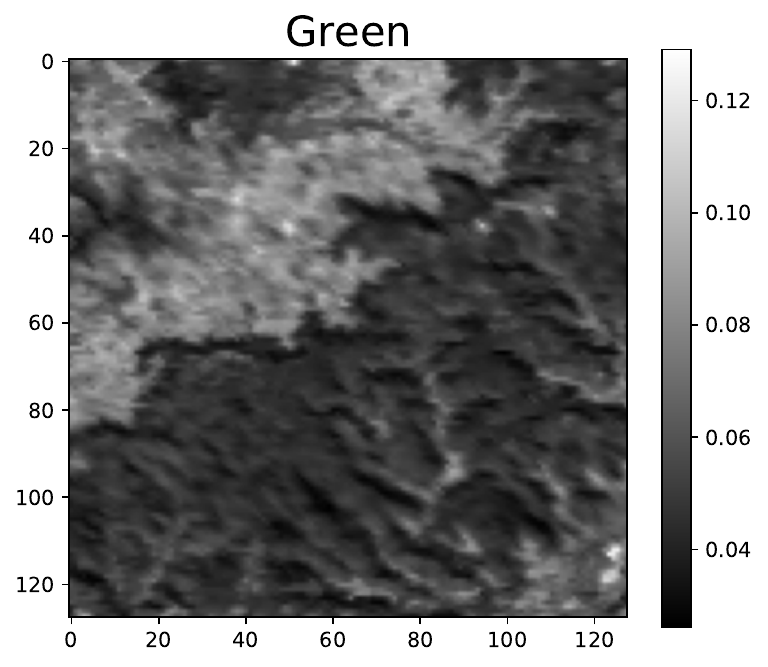}
    \includegraphics[width=0.12\linewidth]{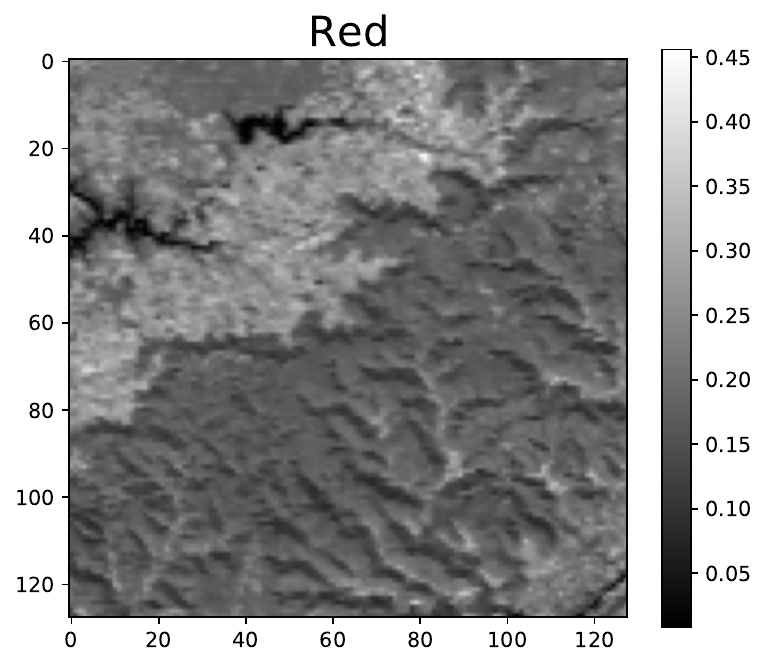}
    \includegraphics[width=0.12\linewidth]{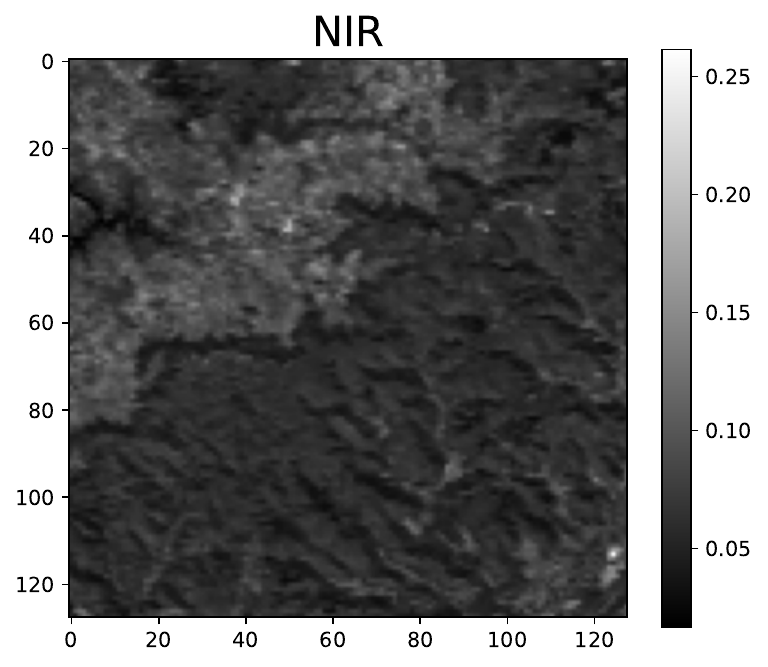}
    \includegraphics[width=0.12\linewidth]{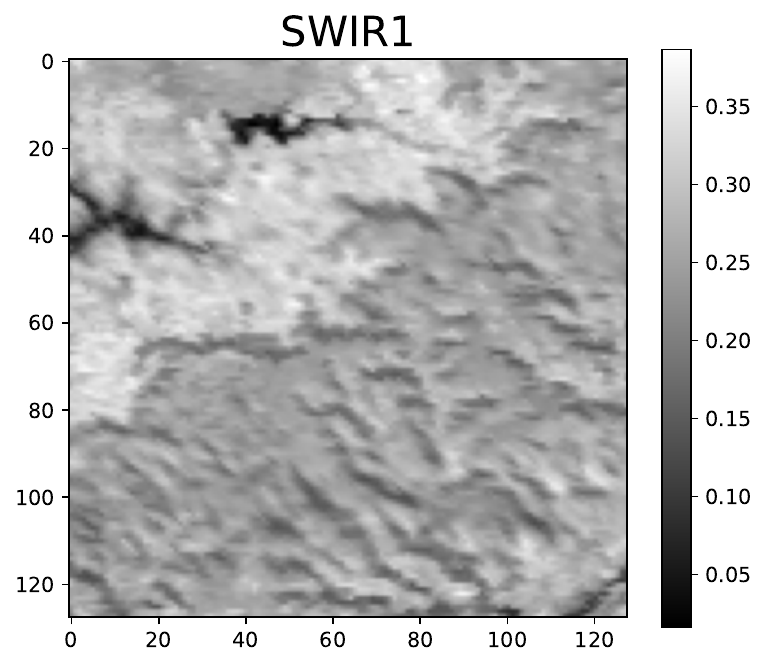}
    \includegraphics[width=0.12\linewidth]{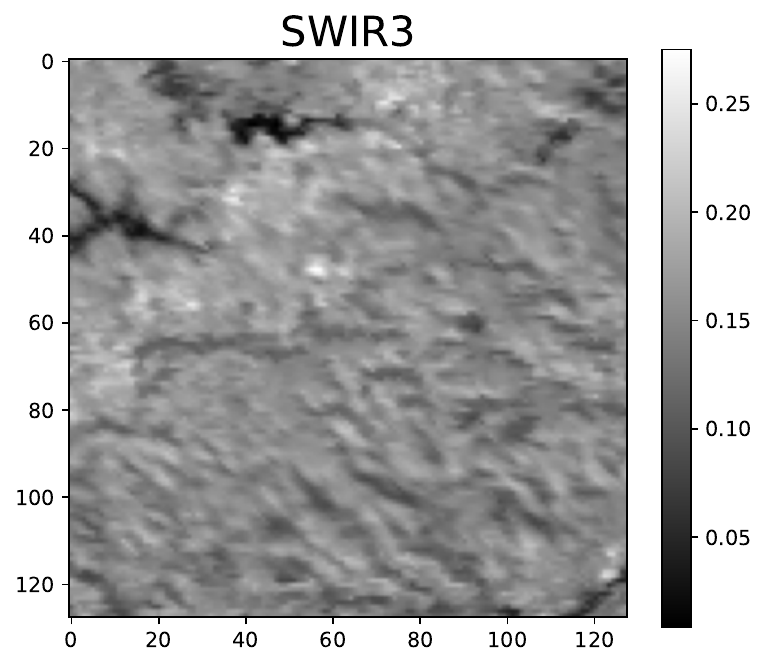}
    \\
    \centering{(a) Multispectral satellite images from MODIS.}
    \\
    \medskip
    {
    \fontsize{6}{7}\selectfont
    \centering
    \begin{minipage}[t]{0.12\linewidth}
      \includegraphics[width=\linewidth]{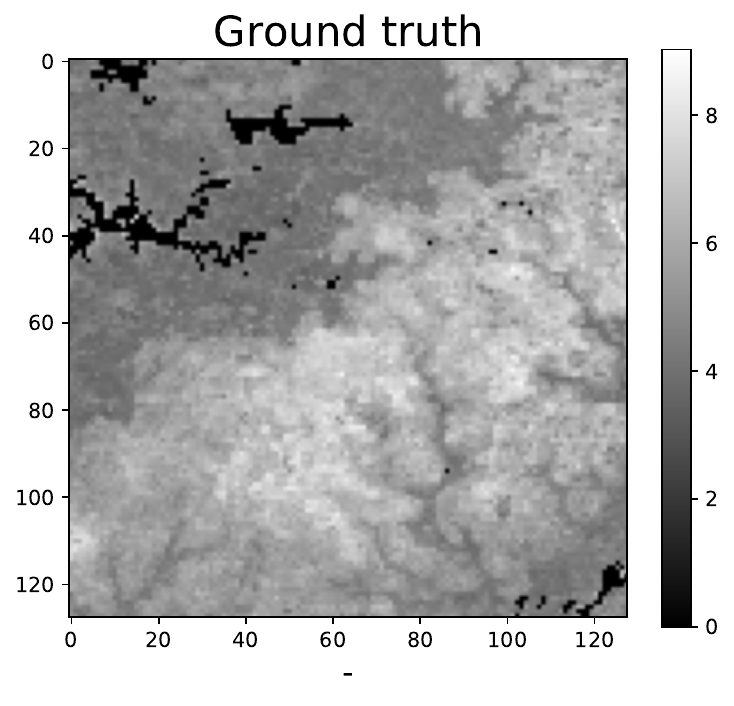} \\
      \centering{Ground truth}
    \end{minipage}
    \vrule
    \begin{minipage}[t]{0.12\linewidth}
      \includegraphics[width=\linewidth]{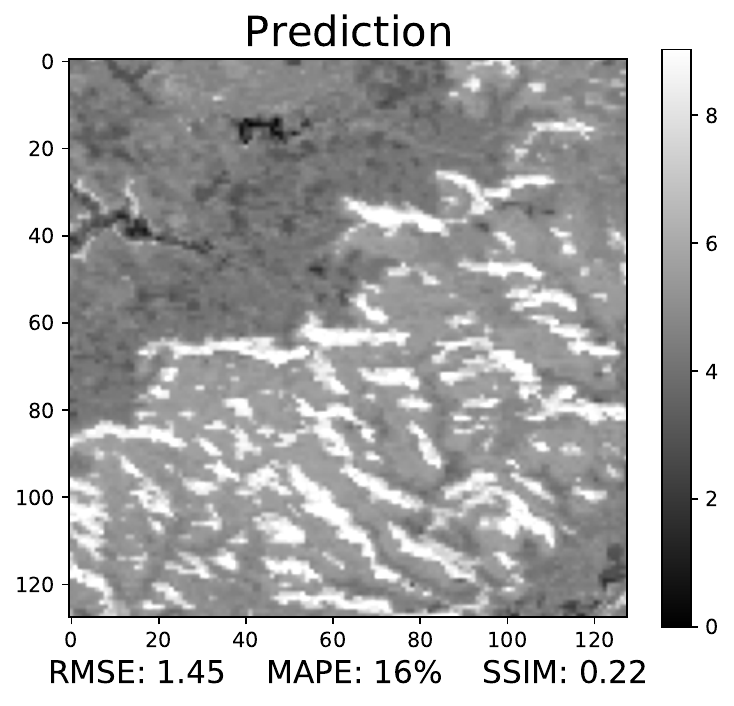} \\
      \centering{Random forest (MAE)}
    \end{minipage}
    \vrule
    \begin{minipage}[t]{0.12\linewidth}
      \includegraphics[width=\linewidth]{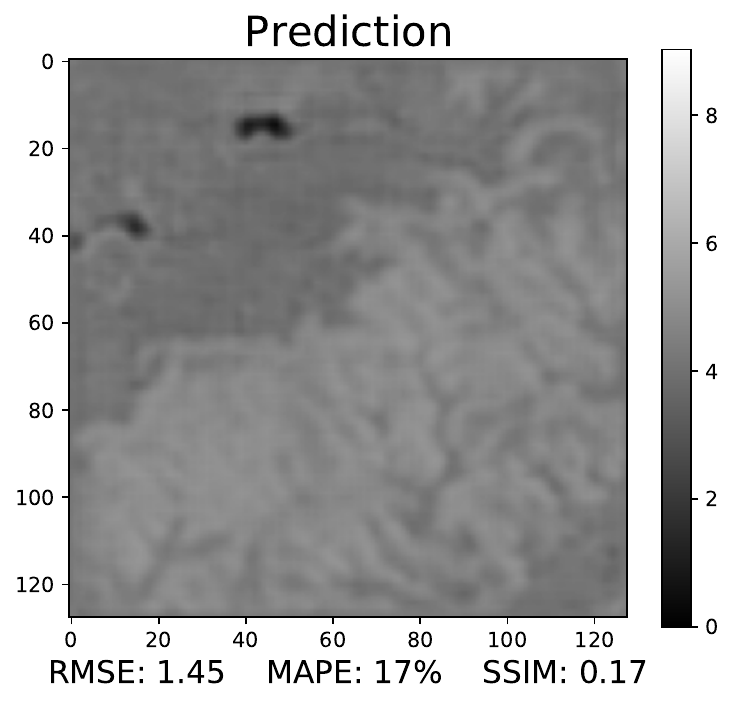} \\
      \centering{V-Net (MAE)}
    \end{minipage}
    \begin{minipage}[t]{0.12\linewidth}
      \includegraphics[width=\linewidth]{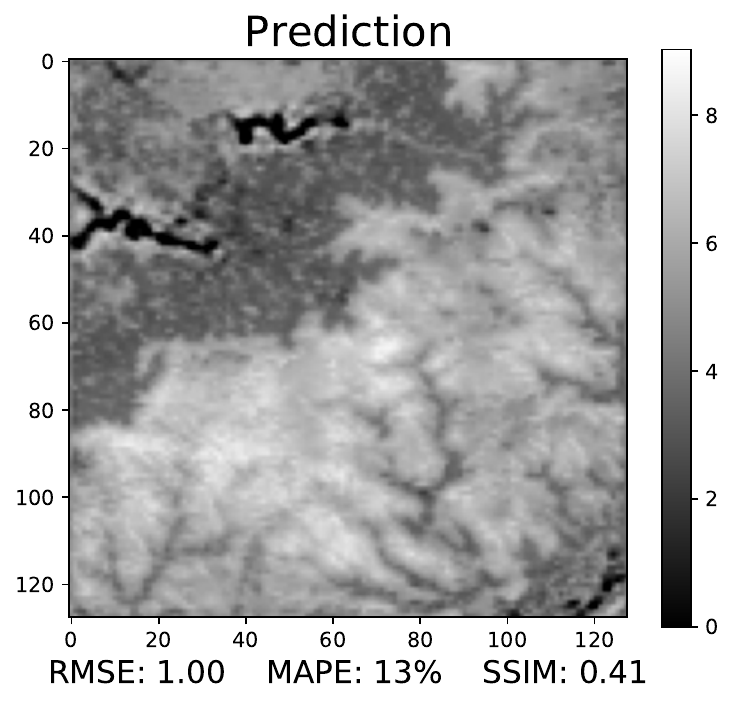} \\
      \centering{V-Net (DSSIM)}
    \end{minipage}
    \begin{minipage}[t]{0.12\linewidth}
      \includegraphics[width=\linewidth]{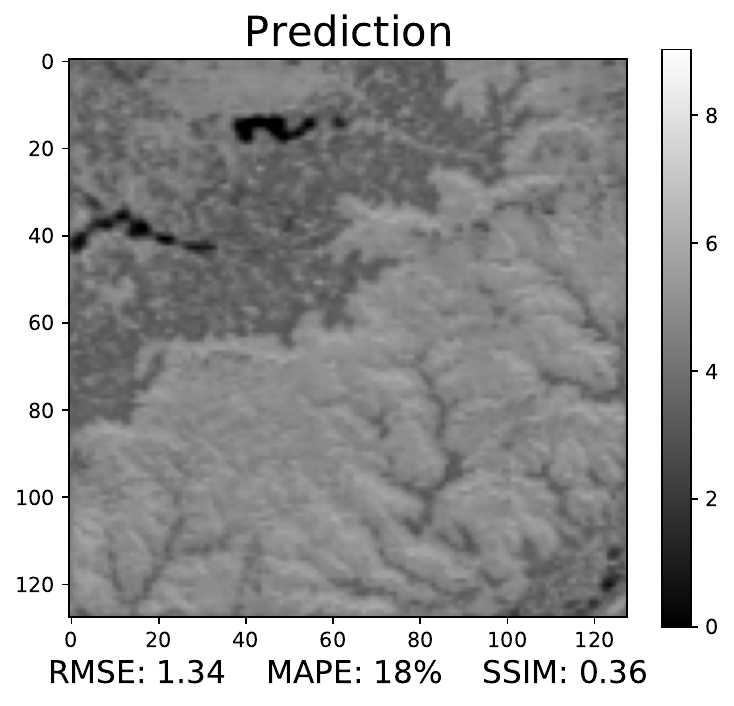} \\
      \centering{V-Net (MAE + DSSIM)}
    \end{minipage}
    }
    \\
    {
    \fontsize{6}{7}\selectfont
    \centering
    \begin{minipage}[t]{0.12\linewidth}
      \includegraphics[width=\linewidth]{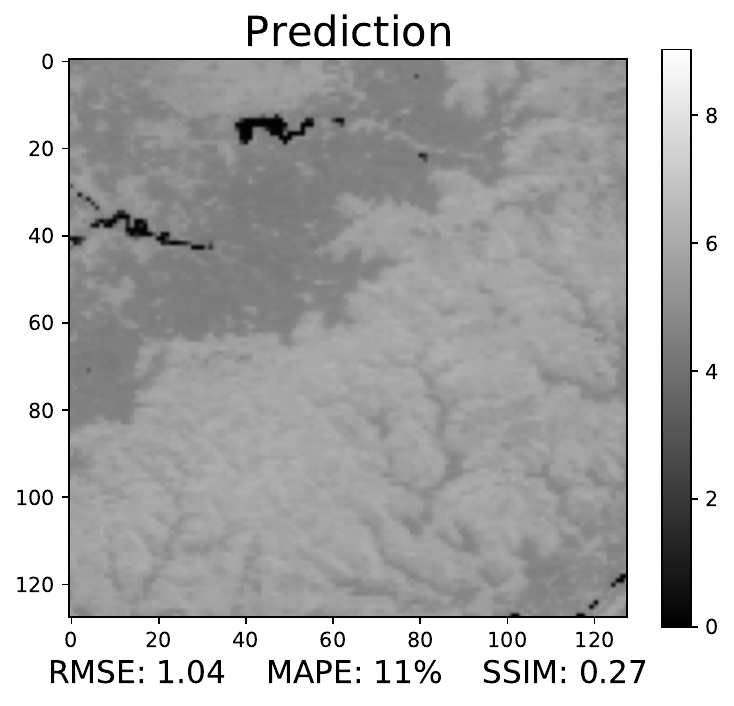} \\
      \centering{FNO (MAE)}
    \end{minipage}
    \begin{minipage}[t]{0.12\linewidth}
      \includegraphics[width=\linewidth]{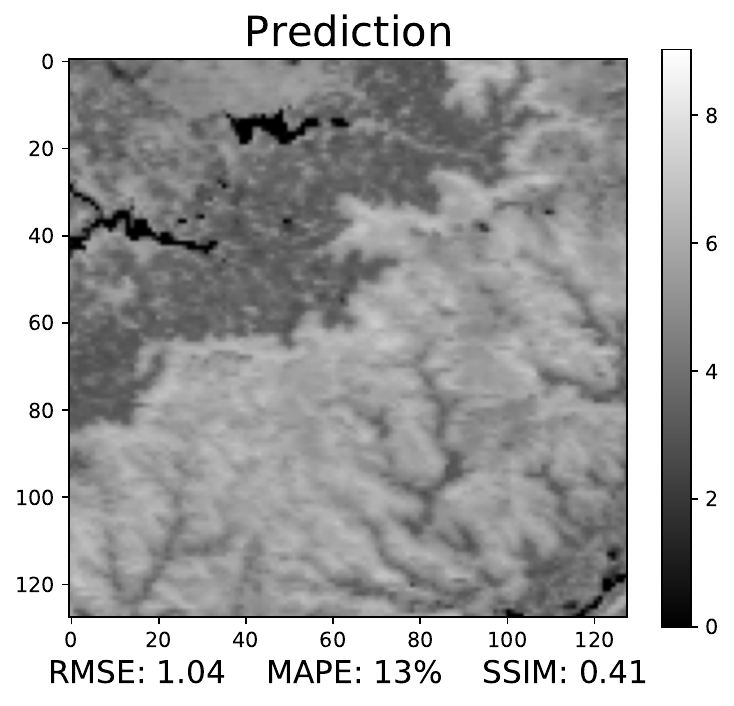} \\
      \centering{FNO (DSSIM)}
    \end{minipage}
    \begin{minipage}[t]{0.12\linewidth}
      \includegraphics[width=\linewidth]{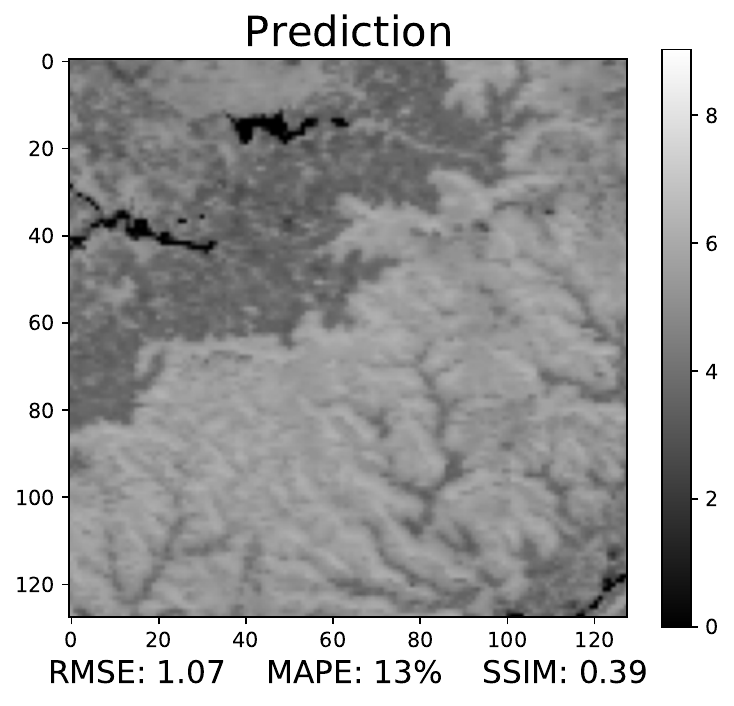} \\
      \centering{FNO (MAE + DSSIM)}
    \end{minipage}
    \vrule
    \begin{minipage}[t]{0.12\linewidth}
      \includegraphics[width=\linewidth]{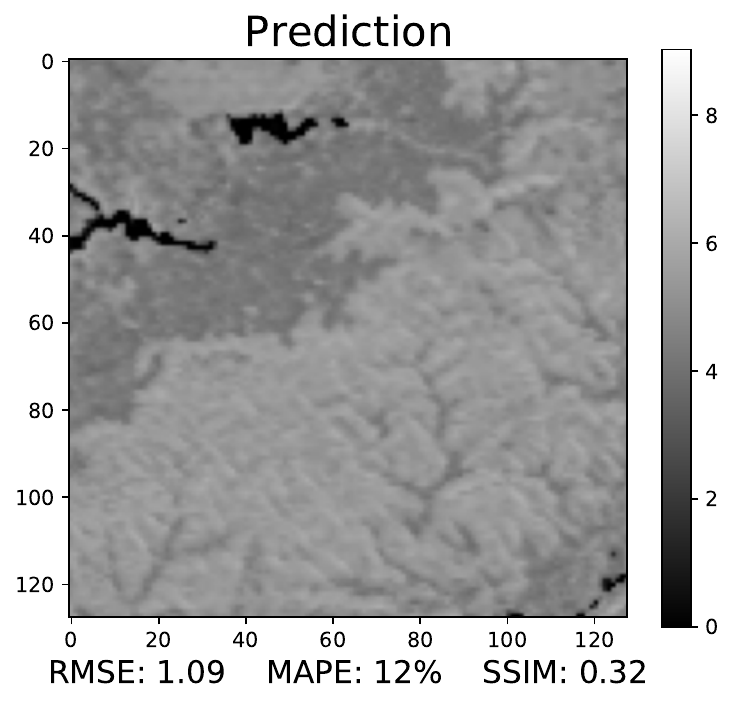} \\
      \centering{FNO-D (MAE)}
    \end{minipage}
    \begin{minipage}[t]{0.12\linewidth}
      \includegraphics[width=\linewidth]{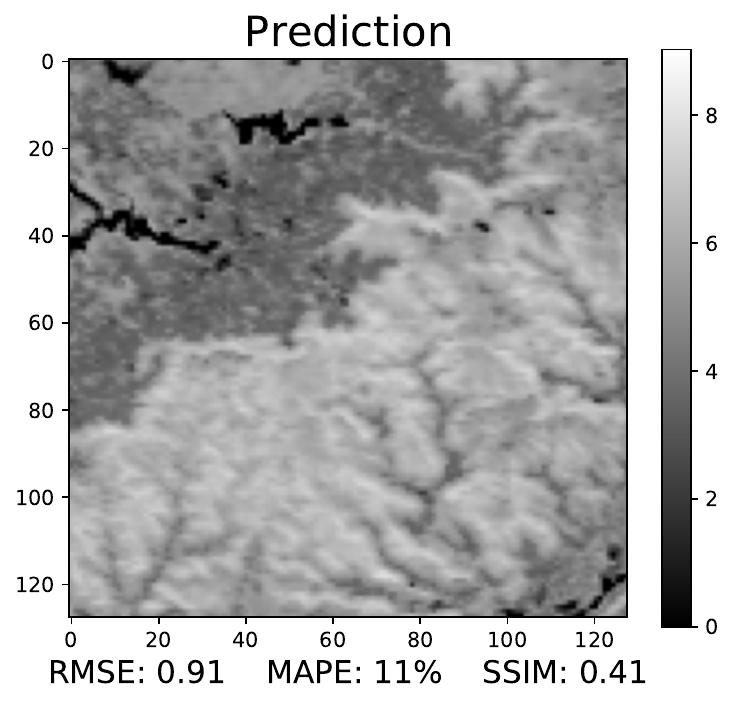} \\
      \centering{FNO-D (DSSIM)}
    \end{minipage}
    \begin{minipage}[t]{0.12\linewidth}
      \includegraphics[width=\linewidth]{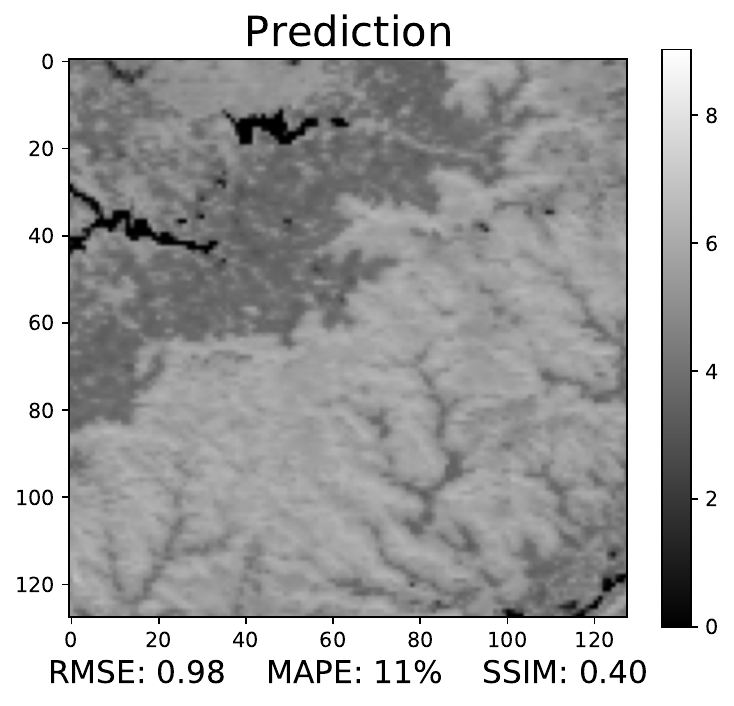} \\
      \centering{FNO-D (MAE + DSSIM)}
    \end{minipage}
    }
    \\
    \medskip
    \centering{(b) Predicted SOC maps of different models and loss functions. FNO-D stands for FNO-DenseNet.}
    \\
    \caption{Visualization of a testing example.}
    \label{fig:visualization}
\end{figure*}
% ---------------------------------------

\section{Experiments}

Our proposed FNO-DenseNet was compared with the original FNO \cite{Conference:Li:ICLR2021:fourier} and a modified V-Net \cite{Conference:Wong:MICCAI2018}. Each architecture was trained with three different combinations of the MAE and DSSIM losses. Furthermore, to compare with the pixel-based approach, a random forest with 10 trees and a maximum depth of 10 was trained with the MAE loss by using each pixel as a sample\footnote{We tried up to 50 trees with different maximum depths.}. To account for the randomness in machine learning models, five models (repeats) were trained for each setting and the average results are reported.

\subsection{Results}

The root mean squared error (RMSE), mean absolute percentage error (MAPE), and SSIM between the predictions and the ground truths on the testing data are reported in Table \ref{table:results}. Comparing among the image-based neural networks, the FNO-DenseNet had the overall best performance and the least number of parameters (64K). Although the original FNO had the second best performance, its number of parameters (34M) was more than 500 times larger than that of the FNO-DenseNet. Comparing among the loss functions, using MAE alone without DSSIM resulted in better MAPE but worse SSIM, and vice versa. When using MAE + DSSIM, we had the best overall results. In contrast, the random forest models had the worst performance. When using only the MAE loss, the random forest and the V-Net had worse SSIM than the FNO and the FNO-DenseNet. This may be due to the global receptive field advantage of the Fourier transform.

Fig. \ref{fig:visualization} shows the visual comparison of a testing example. While the multispectral images had a variety of intensity distributions (Fig. \ref{fig:visualization}(a)), the predictions of the neural networks were similar to the ground truth regardless of the loss functions (Fig. \ref{fig:visualization}(b)). When using only the MAE, regardless of the network architecture, the predictions captured less textural details than those using DSSIM, which is reflected by their relatively low SSIM values. Consistent with Table \ref{table:results}, when using only the MAE, the FNO and FNO-DenseNet had better SSIM values than the random forest and V-Net. The SOC map predicted by the random forest was the most different from the ground truth, which was unable to capture the structural details and some SOC values were overestimated. This is probably caused by the pixel-based nature that does not account for the local dependencies among pixels.

\section{Conclusion}

In this paper, we propose a satellite image-based approach for SOC remote sensing with multispectral satellite data. The experimental results show that by combining the advantages of the FNO and DenseNet, the FNO-DenseNet outperformed the FNO, V-Net, and random forest with the least number of parameters among the neural networks. Furthermore, by using the structural dissimilarity in the loss function, the learned models can provide predictions with structural details like the ground truths. Given these advantages, the proposed framework has the potential to enable measurement, reporting, and verification of SOC across the globe to allow year-to-year tracking of carbon storage and carbon cycle disruptions when carbon sequestration practices are implemented.

\bibliographystyle{IEEEbib}
\bibliography{Ref}

\end{document}